\documentclass[twocolumn,times]{aastex631}

\usepackage{CJK}
\usepackage{times}
\usepackage{newtxtext,newtxmath}
\usepackage{amsmath,mathrsfs,bm}
\usepackage{braket}
\usepackage{subfigure}
\usepackage{color}
\usepackage{xspace}
\usepackage{booktabs}

\usepackage{threeparttable}

\newcommand{\reffg}[1]{Figure~\ref{#1}}
\newcommand{\reftb}[1]{Table~\ref{#1}}
\newcommand{\refeq}[1]{Equation~(\ref{#1})}
\newcommand{\refsc}[1]{Section~\ref{#1}}

\usepackage{etoolbox}
\preto\align{\par\nobreak\noindent}
\expandafter\preto\csname align*\endcsname{\par\nobreak\noindent}
\preto\multline{\par\nobreak\noindent}
\expandafter\preto\csname align*\endcsname{\par\nobreak\noindent}
\preto\flalign{\par\nobreak\noindent}
\expandafter\preto\csname align*\endcsname{\par\nobreak\noindent}
\preto\eqnarray{\par\nobreak\noindent}
\expandafter\preto\csname align*\endcsname{\par\nobreak\noindent}

\def\hiMF{{\textsc{HiMF}}\xspace}
\def\T{{\rm T}}
\def\hi{{\textsc{Hi}}\xspace}
\def\hiim{\textsc{HiIM}\xspace}

\def\apjl{Astrophysical Journal Letters}
\def\aap{Astronomy and Astrophysics}

\def\apjs{The Astrophysical Journal Supplement Series}

\begin{document}
\begin{CJK*}{UTF8}{gbsn}

\title{FAST drift scan survey for \hi intensity mapping: simulation on Bayesian-stacking-based \hi mass function estimation}

\author[0009-0008-7631-7991]{Jiaxin Wang}
\affiliation{Key Laboratory of Cosmology and Astrophysics (Liaoning) \& College of Sciences, Northeastern University, Shenyang 110819, China}

\author[0000-0003-1962-2013]{Yichao Li}
\correspondingauthor{Yichao Li}
\email{liyichao@mail.neu.edu.cn}
\affiliation{Key Laboratory of Cosmology and Astrophysics (Liaoning) \& College of Sciences, Northeastern University, Shenyang 110819, China}

\author[0000-0002-9160-391X]{Hengxing Pan}
\affiliation{National Astronomical Observatories, Chinese Academy of Sciences, Beijing 100101, China}
%\affiliation{Astrophysics, University of Oxford, Denys Wilkinson Building, Keble Road, Oxford OX1 3RH, UK}
%\affiliation{Astrophysics, Department of Physics, University of Oxford, Denys Wilkinson Building, Keble Road, Oxford OX1 3RH, UK}

\author[0000-0001-8075-0909]{Furen Deng}
\affiliation{National Astronomical Observatories, Chinese Academy of Sciences, Beijing 100101, China}
\affiliation{School of Astronomy and Space Science, University of Chinese Academy of Sciences, Beijing 100049, China}
\affiliation{Key Laboratory of Radio Astronomy and Technology, Chinese Academy of Sciences, A20 Datun Road, Chaoyang District, Beijing 100101, China}
\affiliation{Institute of Astronomy, University of Cambridge, Madingley Road, Cambridge, CB3 0HA, UK\\}

\author[0009-0000-6895-9136]{Diyang Liu}
\affiliation{Key Laboratory of Cosmology and Astrophysics (Liaoning) \& College of Sciences, Northeastern University, Shenyang 110819, China}

\author[0009-0006-2521-025X]{Wenxiu Yang}
\affiliation{National Astronomical Observatories, Chinese Academy of Sciences, Beijing 100101, China}
\affiliation{School of Astronomy and Space Science, University of Chinese Academy of Sciences, Beijing 100049, China}

\author[0000-0002-3108-5591]{Wenkai Hu}
\affiliation{Department of Physics \& Astronomy, University of the Western Cape, Cape Town 7535, South Africa}
\affiliation{ARC Centre of Excellence for All Sky Astrophysics in 3 Dimensions (ASTRO 3D), Australia}

\author[0000-0003-0631-568X]{Yougang Wang}
\affiliation{National Astronomical Observatories, Chinese Academy of Sciences, Beijing 100101, China}
\affiliation{School of Astronomy and Space Science, University of Chinese Academy of Sciences, Beijing 100049, China}
\affiliation{Key Laboratory of Radio Astronomy and Technology, Chinese Academy of Sciences, A20 Datun Road, Chaoyang District, Beijing 100101, China}
\affiliation{Key Laboratory of Cosmology and Astrophysics (Liaoning) \& College of Sciences, Northeastern University, Shenyang 110819, China}

\author[0000-0002-6029-1933]{Xin Zhang}
\correspondingauthor{Xin Zhang}
\email{zhangxin@mail.neu.edu.cn}
\affiliation{Key Laboratory of Cosmology and Astrophysics (Liaoning) \& College of Sciences, Northeastern University, Shenyang 110819, China}
\affiliation{National Frontiers Science Center for Industrial Intelligence and Systems Optimization, Northeastern University, Shenyang 110819, China}
\affiliation{Key Laboratory of Data Analytics and Optimization for Smart Industry (Ministry of Education), Northeastern University, Shenyang 110819, China}

\author[0000-0001-6475-8863]{Xuelei Chen}
\correspondingauthor{Xuelei Chen}
\email{xuelei@cosmology.bao.ac.cn}
\affiliation{National Astronomical Observatories, Chinese Academy of Sciences, Beijing 100101, China}
\affiliation{Key Laboratory of Cosmology and Astrophysics (Liaoning) \& College of Sciences, Northeastern University, Shenyang 110819, China}
\affiliation{Key Laboratory of Radio Astronomy and Technology, Chinese Academy of Sciences, A20 Datun Road, Chaoyang District, Beijing 100101, China}
\affiliation{School of Astronomy and Space Science, University of Chinese Academy of Sciences, Beijing 100049, China}

\begin{abstract}

This study investigates the estimation of the neutral hydrogen (\hi) mass function (\hiMF) using a 
Bayesian stacking approach with simulated data for the Five-hundred-meter Aperture Spherical radio Telescope 
(FAST) \hi intensity mapping (\hiim) drift-scan surveys. 
Using data from the IllustrisTNG simulation, we construct \hi sky cubes at redshift $z\sim0.1$
and the corresponding optical galaxy catalogs, simulating FAST observations under various survey strategies, 
including pilot, deep-field, and ultradeep-field surveys.
The \hiMF is measured for distinct galaxy populations -- classified by optical properties 
into red, blue, and bluer galaxies -- and injected with systematic effects such as observational noise
and flux confusion caused by the FAST beam.
The results show that Bayesian stacking significantly enhances \hiMF measurements. 
For red and blue galaxies, the \hiMF can be well constrained with pilot surveys, while deeper surveys are 
required for the bluer galaxy population. 
Our analysis also reveals that sample variance dominates over observational noise, 
emphasizing the importance of wide-field surveys to improve constraints. 
Furthermore, flux confusion shifts the \hiMF toward higher masses, which we address using a transfer 
function for correction. Finally, we explore the effects of intrinsic sample incompleteness and propose a 
framework to quantify its impact.
This work lays the groundwork for future \hiMF studies with FAST \hiim, 
addressing key challenges and enabling robust analyses of \hi content across galaxy populations. 

%We look at the contribution of different galaxy populations to the atomic hydrogen (\hi) mass function (\hiMF) measured by a Bayesian stacking technique directly.
%We simulate the Five-hundred-meter Aperture Spherical radio Telescope (FAST) observation by convolving its beam size and injecting noise into the data, which is released by the IllustrisTNG project, according to different observed repetition of FAST survey at redshift 0.1.
%We classify galaxies into several distinct populations based on their optical properties and compute the \hiMF for each of them by using Multinest program.
%Our main findings are:
%We propose an empirical ratio function in order to explore the influence of beam on sky survey more intuitively.
%Our research is a first step towards the measurement of the \hiMF of different populations of galaxies observed by the FAST survey at higher redshift.

\end{abstract}


\section{Introduction}\label{sec:intro}

Neutral hydrogen (\hi) is the most abundant element in the Universe and plays
an important role in exploring the evolution of the Universe. 
The distribution of \hi content in galaxies is represented by the \hi mass function (\hiMF),
which is defined as the number density of galaxies as the function of \hi mass. 
The measurement of \hiMF across cosmic time is critical for understanding
galaxy formation and evolution.

The \hiMF has been measured within the local Universe via the \hi galaxy surveys, e.g.
the \hi Parkes All-Sky Survey \citep[HIPASS;][]{2001MNRAS.322..486B, 2005MNRAS.359L..30Z} 
and the Arecibo Legacy Fast ALFA survey 
\citep[ALFALFA survey;][]{2005AJ....130.2598G, 2010ApJ...723.1359M, 2018MNRAS.477....2J}, 
which are conducted with the Parkes telescope and the Arecibo telescope, respectively.  A few deeper blind \hi galaxy surveys are currently in progress,
which target to measure the \hiMF to the faint-end and higher redshift
using the newly constructed or upgraded radio telescopes. For instance,  
the Australian Square Kilometre Array Pathfinder \citep[ASKAP;][]{2007PASA...24..174J,5164981}
has carried out the Widefield ASKAP L-band Legacy All-sky Blind Survey 
\citep[WALLABY;][]{2020Ap&SS.365..118K} 
and the Deep Investigation of Neutral Gas Origin \citep[DINGO;][]{2020bugm.conf...18R};
the Netherlands Aperture Tile In Focus \citep[AperTIF;][]{2009pra..confE..10V,2009wska.confE..70O}
which is a phase-array-feed (PAF) upgrade to the Westerbork Synthesis Radio Telescope (WSRT),
is conducting the \hi galaxy imaging survey \citep{2009pra..confE..10V};
and the Meer Karoo Array Telescope \citep[MeerKAT;][]{2009IEEEP..97.1522J}
carries out the MeerKAT International GHz Tiered Extragalactic Exploration (MIGHTEE) survey 
\citep{2016mks..confE...6J} 
and the Looking At the Distant Universe with the MeerKAT Array (LADUMA) survey 
\citep{2012IAUS..284..496H,2016mks..confE...4B}. 

The Five-hundred-meter Aperture Spherical radio Telescope
\citep[FAST,][]{2011IJMPD..20..989N,2016RaSc...51.1060L}
is the largest single-dish radio telescope in the world.
It has an effective aperture of 300 meters and is equipped with a multi-beam feed system and 
low-noise cryogenic receivers \citep{2020raa....20...64j}.
\hi galaxy survey is one of the primary scientific focuses for FAST.
FAST All Sky \hi (FASHI) is one of the FAST \hi survey projects, 
which is proposed to cover the entire sky that can be detected by FAST telescope, 
ranging a declination of $-14^\circ$ to $+66^\circ$ \citep{2024SCPMA..6719511Z}.
The Commensal Radio Astronomy FasT Survey \citep[CRAFTS;][]{2018IMMag..19..112L}
is simultaneously conducting the surveys for pulsar and fast radio burst search, 
the Galactic \hi, and the extragalactic \ galaxies 
\citep{2019SCPMA..6259506Z}.  Recently, ~\cite{2024arXiv241109903M} presented a complete measurement of \hiMF by combining the \hi catalogs from HIPASS, ALFALFA, and FASHI surveys.

\hiMF measurement with galaxy surveys beyond the local Universe is currently sensitivity-limited.
The next-generation telescope, such as the Square Kilometer Array (SKA)
\footnote{\url{https://www.skao.int}} \citep{2015aska.confE..19S,2020PASA...37....7S}, has the potential to
explore the \hiMF across a wide redshift range.
On the other hand, a promising approach to break the flux limit and improving the signal-to-noise 
relies on the stacking technique, which combines the flux measurements from the \hi survey and precise positional information from another survey, particularly galaxy surveys in the optical band. 
By co-adding the spectral line data at the known locations of many galaxies, 
the stacking analysis leads to an enhancement of the signal-to-noise ratio 
\citep{2013MNRAS.433.1398D, 2018MNRAS.473.1879R}.

The stacking technique is also beneficial for \hiMF measurements with \hi intensity mapping (\hiim) surveys,
which measures the total \hi intensity of many galaxies within large voxels 
\citep{2008PhRvL.100i1303C,2008PhRvL.100p1301L,
2008PhRvD..78b3529M,2008PhRvD..78j3511P,2008MNRAS.383..606W,2008MNRAS.383.1195W,
2009astro2010S.234P,2010MNRAS.407..567B,2010ApJ...721..164S,2011ApJ...741...70L,
2012A&A...540A.129A,2013MNRAS.434.1239B}.
The \hiim survey can be quickly carried out and extended to cosmological survey volume,
which is ideal for cosmological investigations and exploring astrophysical processes over a broad cosmic time range.
%and the cosmological investigations
\citep{2015ApJ...798...40X,Zhang:2021yof,Jin:2021pcv,Wu:2021vfz,Wu_2023,2023SCPMA..6670413W,10.1093/mnras/stad2033,Jin:2020hmc,Zhang:2019dyq,Pan:2024xoj}.
The \hiim was first explored with large radio telescope facilities, 
such as the Green Bank Telescope (GBT) and Parkes Telescope
\citep{2010Natur.466..463C,2013ApJ...763L..20M,2018MNRAS.476.3382A,2017MNRAS.464.4938W,
2022MNRAS.510.3495W}.
Currently, there are several ongoing \hiim experiments focusing on the
post-reionization epoch, such as the Tianlai project \citep{2012IJMPS..12..256C,
2020SCPMA..6329862L,2021MNRAS.506.3455W,2022MNRAS.517.4637P,2022RAA....22f5020S}, the
Canadian Hydrogen Intensity Mapping Experiment ~\citep[CHIME,][]{2014SPIE.9145E..22B,2023ApJ...947...16A}.
There are also some \hiim experiments which are under construction, such as 
the Baryonic Acoustic Oscillations from Integrated Neutral Gas Observations
~\citep[BINGO,][]{2013MNRAS.434.1239B} and the
Hydrogen Intensity and Real-Time Analysis experiment ~\citep[HIRAX,][]{2016SPIE.9906E..5XN}.
The \hiim technique is also proposed as the major cosmology project with SKA
and MeerKAT \citep{2015ApJ...803...21B, 2017arXiv170906099S,2021MNRAS.501.4344L,
2021MNRAS.505.3698W,2021MNRAS.505.2039P,2023MNRAS.524.3724C}.
Recently, the MeerKAT \hiim survey reported the cross-correlation power spectrum detection with the
optical galaxy survey using single dish observation mode \citep{2022arXiv220601579C,2024arXiv240721626M,2024arXiv241206750C}
and the auto power spectrum detection at Mpc scales using the MeerKAT interferometric
observation mode \citep{2023arXiv230111943P}.
%%reports the \hiim auto power spectrum detection 
The \hiim auto power spectrum on large scales remains undetected \citep{2013MNRAS.434L..46S}. 

The \hiim can be carried out with FAST drift scan observation \citep{2020MNRAS.493.5854H}.
The initial data analysis pipeline for FAST \hiim drift scan survey was 
investigated with pilot surveys, e.g., 
the FAst neuTral HydrOgen intensity Mapping ExpeRiment \citep[FATHOMER,][]{2023ApJ...954..139L,2024arXiv241202582Z,2024arXiv241103988L} 
and \hi cosmology projects with CRAFTS \citep{2024arXiv241208173Y}.
In this study, we simulate the \hiMF measurements for the FAST \hi drift scan survey
using a Bayesian-stacking-based approach, which was pioneered in the work of 
\citet{2020MNRAS.491.1227P} and further developed in 
\citet{2021MNRAS.508.1897P,2024MNRAS.534..202P} for measuring the \hi scaling relations.

The paper is organized as follows.
In \refsc{sec:data} we briefly introduce the simulation datasets used in this study, 
including the \hi cube and the corresponding optical-selected galaxy sample.
In \refsc{sec:methods} we introduce the Bayesian-stacking-based \hiMF measurement method, 
the FAST observation effect simulation method, and the galaxy classification method. 
In \refsc{sec:results} we show our results and discuss the implications; 
and summarize and conclude in \refsc{sec:conc}.
We use the standard $\Lambda$CDM cosmology with the 
Hubble constant $h$ = 0.6774, total matter density $\Omega_{\rm m}$ = 0.3089 
and dark energy density $\Omega_{\rm \Lambda}$ = 0.6911 \citep{2016A&A...594A..13P}.

\section{Simulation Data}\label{sec:data}

\subsection{The TNG simulation}
The \hi cube is constructed using a snapshot of hydrodynamic simulation.
We use the public hydrodynamic simulation data released by the IllustrisTNG project
\footnote{\url{https://www.tng-project.org}}
\citep[][hereafter TNG]{2019ComAC...6....2N}. 
The TNG project has three different data volumes, i.g. the TNG50, TNG100, and TNG300
corresponding to comoving volume of $35$, $75$, and $205\,h^{-3}{\rm Mpc}^3$, respectively.
In this work, we adopt one snapshot of the TNG100 simulation box at the redshift of $0.1$,  
which has the comoving volume suitable for our FAST \hiim drift scan pilot survey.
The threshold of $2\times 10^8\,{\rm M_{\odot}}$ for gas mass and stellar mass 
is applied to the simulation box following the analysis in the literature \citep{2018ApJS..238...33D}. 
To ensure sample completeness, galaxies with either gas or stellar mass pass the 
threshold are considered reasonable galaxies.

\subsection{\hi sky cube simulation}

\begin{figure*}
  \centering
    \includegraphics[width=\textwidth]{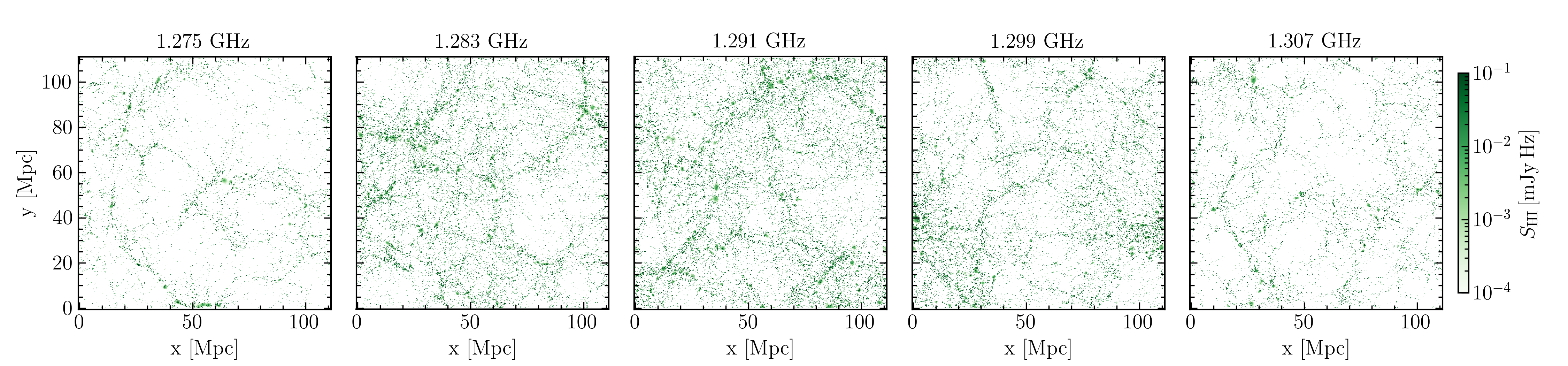}
\caption{ The \hi flux distribution of simulated sky map at different frequencies
integrated across a frequency range of $8.0\,{\rm MHz}$. }\label{fig:sim_map}
\end{figure*}

The total neutral gas mass, including both the \hi and H$_2$, is provided by the TNG simulation.
The TNG simulation utilizes different neutral gas models for the high-density and low-density regions,
i.e., in the high-density regions, the neutral fraction is calculated considering the
effects of cooling, heating, and ionization, while in the low-density regions, the gas
is assumed to be in ionization equilibrium with a self-shielding effect against 
the time-evolving ultraviolet(UV)/X-ray background. 
In addition, the heating and ionization effect of stellar and active galactic nucleus (AGN) feedback 
is also coupled in the calculation. For a more detailed discussion, refer to the work of
\citet{2018ApJS..238...33D}.
The \hi fraction is then determined according to the H$_2$ abundance following the 
model from \citet{2011ApJ...728...88G}. For more details of the \hi to H$_2$ fraction, 
refer to the work of \citet{2022MNRAS.515.5894D}.

The TNG simulation box is then gridded into a sky cube with a pixel area of 
$20\times 20\,{\rm kpc}^2$
and frequency resolution of $0.1\,{\rm MHz}$. We use flat sky approximation with 
$z-$axis of the TNG simulation box as the line-of-sight (LoS) direction and 
set the $z-$axis center as the comoving distance of redshift $0.1$.
The projected sky area is $\sim 210\,\deg^2$, which is comparable to the currently 
FAST \hiim drift scan pilot survey.
The comoving distance of each particle is then converted to observer-frame frequency 
under the assumption of the $\Lambda$CDM cosmology model \citep{2016A&A...594A..13P}. 
In addition, the LoS projection of the peculiar velocity for each particle is also considered
to form a realistic galaxy \hi emission line profile. 
The frequency of particles in the observer frame is calculated via, 
\begin{align}
    \nu_{\rm obs} =\frac{\nu_0}{1 + z + v_{\rm z}/c},
\end{align}
where $\nu_{\rm obs}$ is the particle's frequency at observer frame, 
$z$ is the particle's cosmological redshift,
$\nu_0 = 1.42\,{\rm GHz}$ is the \hi rest-frame frequency, $c$ is the speed of light, and
$v_{\rm z}$ is the LoS projection of the particle's peculiar velocity.

The rest-frame velocity integrated \hi flux of each particle is then calculated via \citep{2017PASA...34...52M},
\begin{align}\label{eq:mass2flux}
\bigg(\frac{S_\hi}{{{\rm Jy\,Hz}}}\bigg) \simeq
\frac{1}{49.7}
\bigg(\frac{m_{\hi}}{h^{-2}{\rm M}_{\odot}}\bigg) 
\bigg(\frac{d_{\rm L}}{h^{-1}{\rm Mpc}}\bigg)^{-2},
\end{align}
where $m_\hi$ and $d_{\rm L}$ are the \hi mass and luminosity distance
of the particles, respectively.
The flux of particles involved in the same pixel of the sky cube is then integrated.
The pixel size and frequency resolution are fine enough to resolve both the galaxy image
and the \hi profile. 
The frequency slices of the simulated sky cube are shown in \reffg{fig:sim_map}.

\subsection{The optical galaxy survey catalog}\label{sec:optcat}

The optical galaxy survey catalog, which is required as the proxy for \hi-stacking analysis, 
is also simulated using the corresponding galaxy sample of the TNG simulation.
The survey area and available frequency range for FAST \hi survey 
coincide with the galaxy catalog of Sloan Digital Sky Survey
\footnote{\url{https://www.sdss.org}} (SDSS). 
In particular, the main galaxy sample (MGS) of the SDSS Data Release 7 \citep[DR7;][]{Abazajian_2009}
covers the redshift range $0<z<0.2$, which corresponds to the high-frequency band of the FAST \hi survey.
The MGS galaxy catalog occupies a footprint of $\sim6800\,\deg^2$.
The MGS could be approximately constructed using magnitude cut,
\begin{align}
14.5 < r_{\rm pet} < 17.6,
\end{align}
where $r_{\rm pet} $ is the extinction-corrected $r$-band Petrosian magnitude
\citep{2015MNRAS.449..835R}.

The photometry measurements of the TNG galaxy samples are simulated 
using method similar to that of \citet{2022MNRAS.515.5894D}, which 
adopte the synthetic stellar photometry model of \citet{2018MNRAS.475..624N}.
We use model B in \citet{2018MNRAS.475..624N} and calculate the spectrum 
of each particle via the stellar population synthesis code
\citep{2009ApJ...699..486C,2010ApJ...712..833C}.
The effect of dust obscuration is modeled using the dust component model
of \citet{2000ApJ...539..718C}.
The galaxy spectrum is then convolved with SDSS instrumental bandpass to 
generate the corresponding apparent magnitude in the observational frame.
Eventually, we apply a three-dimensional radial constraint of no more than 
$30\,{\rm kpc}$ from the subhalo center, which approximates a Petrosian aperture 
and yields an apparent magnitude similar to that utilized in MGS galaxy selection.  
For more details of the galaxy apparent magnitude simulation, refer to 
the work of \citet{2022MNRAS.515.5894D}.

We apply the same apparent magnitude cut as the SDSS MGS galaxy catalog 
to the simulated TNG galaxy sample and restrict the redshift range between 
$0.02 < z < 0.2$ to match the frequency range of the simulated \hi sky cube.

In addition, following the data partition in the real observation data analysis
\citep{2020MNRAS.494.2664D}, 
we split the simulated MGS-like galaxy samples into three
subsamples, i.e. the red, blue, and bluer galaxies. 
The detailed subsample classification method is presented in \refsc{sec:galaxypop}.
The number density of galaxies in MGS-like sample, 
as well as the three subsamples, are listed in \reftb{tab:numdensity}. 
We also list the galaxy number density of the same categories from
the real observational data of \citet{2020MNRAS.494.2664D} as the reference.

\begin{table}
\scriptsize
\centering
\caption{
Comparing the galaxy number density of the catalogs generated by the 
TNG simulation with the observational catalog utilized in \citet{2020MNRAS.494.2664D}.
The galaxy number densities are in units of $h^{3}{\rm Mpc}^{-3}$.
}\label{tab:numdensity}
\begin{tabular}{lcc}
\hline\hline
& Observational catalog & TNG simulation \\ \hline
Total & $1.23\times10^{-2}$ & $0.97 \times 10^{-2}$ \\
Red   & $1.78\times10^{-3}$ & $2.50 \times 10^{-3}$ \\
Blue  & $6.63\times10^{-3}$ & $6.53 \times 10^{-3}$ \\
Bluer & $6.47\times10^{-4}$ & $6.69 \times 10^{-4}$ \\ \hline
\hline
\end{tabular}
\end{table}

\section{Methods}\label{sec:methods}

\subsection{\hi mass function}\label{sec:himf}

The \hiMF represents the intrinsic number density of galaxies as a function of their \hi mass. 
By definition the \hiMF can be expressed as \citep{2002ApJ...567..247R}
\begin{equation}\label{eq:himf}
    \Phi(M_\hi)=\frac{{\rm d}N_{\rm gal}}{{\rm d}V\,{\rm d\,log_{10}}(M_{\rm \hi})},
\end{equation}
where ${\rm d}N_{\rm gal}$ represents the mean galaxy number in a comoving volume ${\rm d}V$, where the galaxy \hi mass falls inside a logarithmic mass bin centered on $M_\hi$.In this work, we estimate the \hiMF via \refeq{eq:himf} with the full knowledge of the galaxy 
\hi mass of the TNG simulation. 
It reflects the \hiMF measurement via a highly sensitive \hi galaxy survey and is denoted as $\Phi_{\rm g}(M_\hi)$.
We employ $\Phi_{\rm g}(M_\hi)$ as the proxy for the true \hiMF and compare it to the \hiMF estimated via the Bayesian stacking of the \hi fluxes in the following analysis when considering different galaxy selection criteria.

The \hiMF at redshift $z$ can be fitted by the Schechter function \citep{2005MNRAS.359L..30Z, Martin_2010},
\begin{align}
    \Phi (M_{\rm \hi},z)=\ln(10) \, \Phi_*\bigg(\frac{M_\hi}{M_*}\bigg)^{\alpha +1}
    e^{-\frac{M_\hi}{M_*}}\bigg(1+z\bigg)^{\beta},
\end{align}
where $\Phi_*$ represents the normalization constant, $M_*$ characters the 'knee' mass of the \hiMF,
$\alpha + 1$ is the faint-end slope, and $\beta$ represents the redshift evolution. 
The volume-weighted mean \hiMF over a redshift range from $z_1$ to $z_2$ is defined as,
\begin{align}\label{eq:himfz}
\Phi(M_\hi, z_1, z_2) = \frac{1}{\tilde{V}} \int_{z_1}^{z_2} \frac{{\rm d}V}{{\rm d}z} 
\Phi(M_\hi, z) {\rm d}z,
\end{align}
where ${\rm d}V/{\rm d}z$ is the differential comoving volume and 
$\tilde{V} = \int_{z_1}^{z_2} {\rm d}V$ is the normalization factor.
In this work, the redshift evolution of the \hiMF is ignored, and parameter $\beta$ is fixed to $\beta=0$.
We use $\Phi_*$, $\alpha$, and $M_*$ as fitting parameters.

\subsection{Bayesian stacking measurements of \hiMF}

The key innovation of the Bayesian stacking method is directly estimating the 
\hiMF using the galaxy number density function binned by measured flux. This method was explored in the previous work of \citet{2020MNRAS.491.1227P}. We briefly summarize the basic procedures. We refer the interested readers to \cite{2020MNRAS.491.1227P} for more details.

The first step of the Bayesian stacking is to find a catalog of galaxies, which provide a proxy of galaxy locations. For the \hiMF measurement, typically an optical galaxy survey catalog can be used. In this work, as described in \refsc{sec:optcat}, we adopt a simulated galaxy catalog that mirrors the selection criteria of the SDSS main galaxy sample.

We then extract the flux measurements, $S_{\rm m}$, from the pixels corresponding to the positions of optically identified galaxies. 
The galaxy number counts, $k_i$, for the $i$-th flux measurement interval, 
$\Delta S_i = [S_i, S_i + \delta S]$, are then determined from the radio survey. 
Instead of using uniform flux intervals, we employ the Bayesian Block method as proposed by \citet{2020MNRAS.491.1227P}. This method dynamically determines optimal bin edges for the dataset using Bayesian statistics \citep{2013ApJ...764..167S,2013A&A...558A..33A,2018AJ....156..123A}.
The Bayesian Block approach adaptively adjusts bin sizes to emphasize regions 
with significant variation while minimizing noise in sparser regions, 
resulting in more robust statistical constraints.

The measurement obtained from the above stacking analysis is the source count 
within a flux interval, $\Psi(S_\hi) = {\rm d}N/{\rm d}S_\hi$. 
If the noise can be ignored, the source count within the 
flux interval is related to the \hiMF by,
\begin{align}
\Psi(S_\hi) = 
\frac{{\rm d}N}{{\rm d}S_\hi} = \frac{1}{\ln (10)}\frac{\Phi(M_\hi, z_1, z_2)}{M_\hi}\frac{{\rm d}M_\hi}{{\rm d}S_\hi},
\end{align}
where ${\rm d}M_\hi/{\rm d}S_\hi = 2.35\times10^5 d_{\rm L}^2/\left(1+z\right)$ is derived 
from the conversation between galaxy \hi mass and its flux \citep{2017PASA...34...52M},
$\Phi(M_\hi, z_1, z_2)$ is the \hi mass function defined with \refeq{eq:himfz}.
The posterior probability distribution of $\Psi(S_\hi)$, 
as well as the fitting parameters involved in the \hiMF model is propagated
to the posterior probability distribution of the measurements via the Bayes' theorem, 
\begin{align}
P\left(\Psi(S_\hi) | k_i \right) \propto& P\left(k_i | \Psi(S_\hi)\right) 
\end{align}

We assume $P\left(k_i | \Psi(S_\hi)\right)$ obeys the Poisson distribution, and the 
probability function is expressed as,
\begin{align}\label{eq:pki}
P\left(k_i | \Psi(S_\hi)\right) = \frac{\lambda_i^{k_i}}{k_i !} e^{-\lambda_i},
\end{align}
where $\lambda_i$ is the expectation value of the source number in the flux interval.

In actual observation, noise introduces 
additional flux and affects the source number counts $k_i$. Some galaxies may have weak \hi fluxes that fall below the noise variance. The expected value of the source number, $\lambda_i$, in the $i$-th flux interval is estimated by convolving the $\Psi(S_{\hi})$ distribution with the noise distribution function, 
\begin{align}
\lambda_i = \int {\rm d}S_\hi \left( \Psi(S_\hi) \int_{\Delta S_i} {\rm d}S_{\rm m} 
P_{\rm N}(S_{\rm m} - S_\hi) \right).
\end{align}
Given the measured flux $S_{\rm m}$ and \hi flux $S_\hi$, $P_{N}(S_{\rm m} - S_\hi)$ represents the distribution of flux difference raised by the noise. We model the noise distribution as a zero-centered Gaussian with a variance of $\sigma_{\rm noise}^2$,
\begin{align}
P_{\rm N}(S_{\rm m} - S_\hi) = \frac{1}{\sigma_{\rm noise}\sqrt{2\pi}} 
\exp\left(-\frac{\left(S_{\rm m} - S_\hi\right)^2}{2\sigma_{\rm noise}^2}\right).
\end{align}

Including the noise probability function, $\lambda_i$ is expressed as,
\begin{align}
\lambda_i= \int{\rm d}S_\hi \frac{\Psi(S_\hi)}{2} \left[ 
 {\rm erf}\left(\frac{S_i + \delta S - S_\hi}{\sqrt{2}\sigma_{\rm niose}}\right)
-{\rm erf}\left(\frac{S_i - S_\hi}{\sqrt{2}\sigma_{\rm noise}}\right)\right],
\end{align}
where the integration interval covers the full \hi flux range.

The likelihood function is constructed by combining the posterior distribution functions, as defined in \refeq{eq:pki}, across all flux intervals:
\begin{align}
\mathcal{L} \propto \prod_{i} P(k_i | \Psi(S_{\hi})) = \prod_{i} \frac{\lambda_i^{k_i}}{k_i!} e^{-\lambda_i},
\end{align}
where $k_i$ represents the observables extracted from radio measurements. The fitting parameters involved in $\lambda_i$ are constrained by maximizing the likelihood function. To explore the parameter space and posterior distributions, we use the multimodal 
nested sampling algorithm, 
{\sc MultiNest}\footnote{\url{https://github.com/JohannesBuchner/PyMultiNest}} \citep{2009MNRAS.398.1601F}.

\subsection{FAST noise level}

We consider the \hiMF measurements uncertainty raised by the observation noise.  
Assuming the noise to be Gaussian, the noise level for a dual-polarization single-beam telescope 
observation can be well approximated by,
\begin{equation}
\sigma _{\rm noise}= \sqrt{2} \frac{k_{\rm B} T_{\rm sys}} {A_{\rm eff}}\frac{1}{\sqrt{\Delta v t}},
\end{equation}
where $k_{\rm B}$ is the Boltzmann constant, $\Delta v$ is the frequency interval for a channel, 
$A_{\rm eff}$ $\approx$ 50000 $\,{\rm m}^2$ is the effective aperture area, 
and $t$ is the total integration time. In this work, we assume the FAST \hiim survey 
is carried out with the drift scan observation mode. The integration time for one transit across 
one FAST beam is approximated by \citep{2020MNRAS.493.5854H},
\begin{align}
t_{\rm FWHM} = \theta_{\rm B} / (\omega_{\rm E} \cos \delta),
\end{align}
where $\theta_{\rm B}$ represents the full-width-half-maximum beam width,
\begin{align}\label{eq:beam}
\theta_{\rm B} = 2.94(1 + z)\,{\rm arcmin} ,
\end{align}
$\omega_{\rm E} \simeq 0.25\,{\rm arcmin}\,{\rm s}^{-1}$ is the angular velocity of the Earth rotation,
and $\delta$ is the Declination angle of the pointing direction.
Assuming the drift scan pointing at the Zenith, where $\delta \simeq 26\, \deg$, 
the integration time is about $14.4\,{\rm s}$ per beam at the redshift $z \sim 0.1$
close to the mean redshift of the MGS-like galaxy samples. 

During our FAST \hi drift scan {\it pilot survey}, the 19-feed array is employed and 
rotated by $23.4^\circ$ to optimize the sky coverage. 
Taking into account the overlapping between the 19 beams, the sky can be scanned twice, 
i.e. $t_{\rm pilot} = 29\,{\rm s}$ per drift-scan observation.
A {\it deep-field survey} that repeats the drift scan four times on the same $\sim 210\,\deg^2$ 
sky area will be examined in this work for FAST \hiim observations.
An approximation for the integration time is $t_{\rm deep} = 115\,{\rm s}$.
We'll also explore an {\it ultradeep-field survey}, which involves doing the drift scan on the 
same $\sim 210\,\deg^2$ sky area eight times.
Approximately $t_{\rm ultra} = 230\,{\rm s}$ is the integration time.
The \hiim drift scan surveys considered in this work are summarized in \reftb{tab:surveys}.

\begin{table}
\scriptsize
\begin{center}
\caption{The FAST \hiim drift scan surveys considered in this work.}\label{tab:surveys}
\hspace*{-1.0cm}
\begin{tabular}{lccc} \hline\hline
              & Pilot survey   & Deep-field survey   & Ultradeep-field survey \\ \hline
$S_{\rm area}$$^{\S}$ & $210\,\deg^2$  & $210\,\deg^2$ & $210\,\deg^2$  \\ 
$n_{\rm repeat}$$^{\dag}$  & $1$            & $4$           & $8$ \\
$t_{\rm pix}$$^{\ddag}$ & $29\,{\rm s}$  & $115\,{\rm s}$ & $230\,{\rm s}$ \\ 
\hline\hline
\end{tabular}
\end{center}
{\footnotesize
\noindent
$^{\S}$ The survey area \\
$^{\dag}$ The repeating times of the drift-scan observation \\
$^{\ddag}$ The integration time per pixel\\
}
\end{table}

$T_{\rm sys} = T_{\rm rec} + T_{\rm sky}$ is the system temperature,
where $T_{\rm rec} $ is the receiver temperature %\citep{2023ApJ...954..139L},
and the sky temperature is modeled as \citep{2020MNRAS.493.5854H},
\begin{equation}
    T_{\rm sky}=T_{\rm CMB} + 25.2\,{\rm K} \times \left ( 0.408/\nu_{\rm GHz} \right )^{2.75},
\end{equation}
where $T_{\rm CMB} = 2.73\,{\rm K}$ is the mean temperature of the cosmic microwave background (CMB).
The $T_{\rm sys} \sim 24\,{\rm K}$ at the target frequencies.  
We adopt a slightly higher system temperature, following the system temperature estimation with FAST real data, 
to account for potentially extra noise contributions \citep{2023ApJ...954..139L}.

\subsection{Correction for the flux confusion effect}

The FAST beam size is approximate via \refeq{eq:beam}, corresponding to the transverse comoving distance 
$\sim 260\,h^{-1}\,{\rm kpc}$ at redshift $z \simeq 0.1$.
%\red{please double check the values here}. 
The flux measurements could be confused since multiple galaxies may not be resolved \citep{2020MNRAS.493.5854H}.
Such confusion effect should be corrected for \hiMF estimation. 

The FAST beam profile is modeled via a Gaussian function, 
\begin{align}
B(\theta) = \exp\left(-\frac{1}{2}\frac{\theta^2}{\sigma_{\rm B}^2}\right),
\end{align}
where $\sigma_{\rm B} = \theta_{\rm B} / (2\sqrt{2\ln 2})$ is frequency dependent
and $\theta_{\rm B}$ is the full-width-half-maximum defined in \refeq{eq:beam}.
The simulated sky cube is convolved with the beam function frequency-by-frequency.
The Bayesian stacking analysis for the \hiMF estimation is then applied to the beam-convolved sky cube.
The transfer function for confusion correction is defined as the 
the ratio of the beam-convolved \hiMF, $\Phi_{\rm B}$, to the reference \hiMF, $\Phi_{\rm ref}$,
\begin{align}\label{eq:Tcc}
T_{\rm cc} = \Phi_{\rm B} / \Phi_{\rm ref},
\end{align}
where $\Phi_{\rm ref}$ is estimated by Bayesian-stacking the \hi flux extracted 
from the simulated galaxy catalog.

In this analysis, the flux leakage from the side lobe, as well as the shape variation 
between the 19 beams, are negligible compared to the flux measurement confusion raised from
the beam smearing effect. However, the imperfect beam profile might complicate the 
foreground subtraction for \hiim survey and result in foreground 
residual contamination in the actual observations. 
We also ignore the foreground residual contamination and assume it can be further 
eliminated in future analysis \citep{2022ApJ...934...83N,2023MNRAS.525.5278G}.

\subsection{Galaxy population}\label{sec:galaxypop}

\begin{figure}
  \centering                           
  \includegraphics[width=0.5\textwidth]{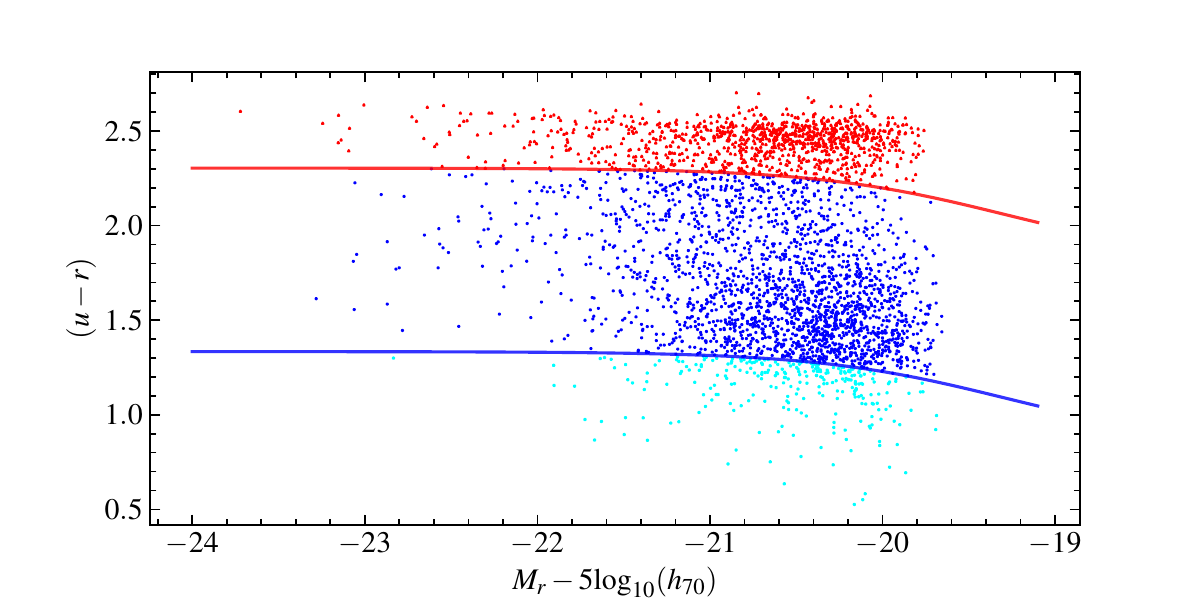}
\caption{
The color-magnitude space distribution of the galaxies from the simulated MGS catalog.
The red curve shows the red/blue galaxy separator locus, which is defined
with \refeq{eq:cur}, and the blue curve shows the separator for the blue/bluer galaxy.
The red, blue, and bluer galaxies from the simulated MGS catalog are shown with the red, cyan, and blue
scatter points, respectively.
\label{fig:colorMr}}
\end{figure}

Understanding the relationship between the galaxy population and the distribution of 
\hi galaxies is essential to comprehending the evolution history of the galaxy. 
These investigations are now conducted using several \hi galaxy surveys, which are restricted to 
the local Universe. For instance, based on the MGS catalog and ALFALFA \hi galaxy survey,  
previous studies \citep[e.g.][]{2020MNRAS.494.2664D} have explored the relation between the \hiMF
and the galaxies with different colors.
Future \hiim surveys could expand this study with the assistance of the stacking analysis.
With the stacking analysis, galaxy population classification relies on the galaxy property 
of an optically selected galaxy catalog.
In this work, we focus on the \hiMF dependency on the galaxy population classified 
from the color-magnitude plane.

The SDSS MGS includes galaxies with their properties spanning a significant range in the 
color-magnitude plane, where the color is the $u-r$ color and magnitude is the 
SDSS $r$-band absolute magnitude. 
The $u-r$ color has been shown to be an optimal separator for the red and blue galaxies
\citep{2001AJ....122.1861S}, which is defined as a locus in the color-magnitude space \citep{2004ApJ...600..681B},
\begin{equation}\label{eq:cur}
    c_{\rm ur}(M_r)= 2.06-0.244 \tanh\left(\frac{M_r+20.07}{1.09}\right).
\end{equation}
The red/blue galaxy separator locus is shown as the red curve in \reffg{fig:colorMr}.

Following the processing in \citet{2020MNRAS.494.2664D}, we further split $23\%$ blue
galaxies, which have their $u-r$ color approaching the lower end of the
galaxy distribution in the color-magnitude space.
Such galaxies are referring as bluer galaxies and the separator locus is chosen as
a curve parallel to the red/blue separator. The blue/bluer separator locus is shown
with the blue curve in \reffg{fig:colorMr}.

\section{Results and Discussion}\label{sec:results}

\subsection{Bayesian-stacking with MGS-like catalog}

\begin{figure}
  \centering
  \includegraphics[width=0.5\textwidth]{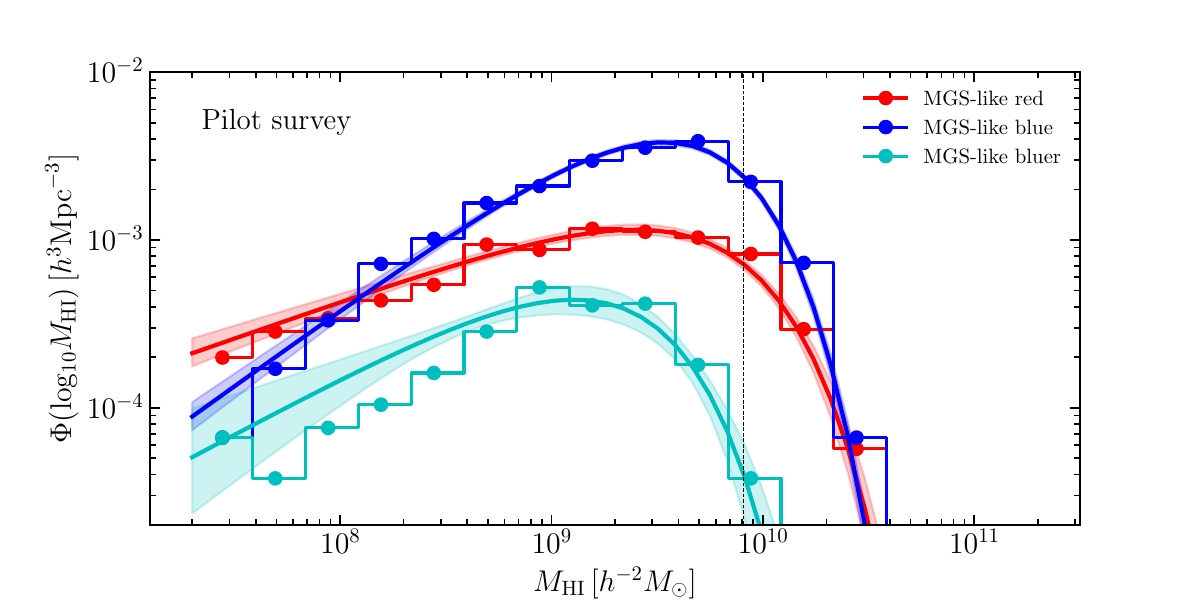}
  \includegraphics[width=0.5\textwidth]{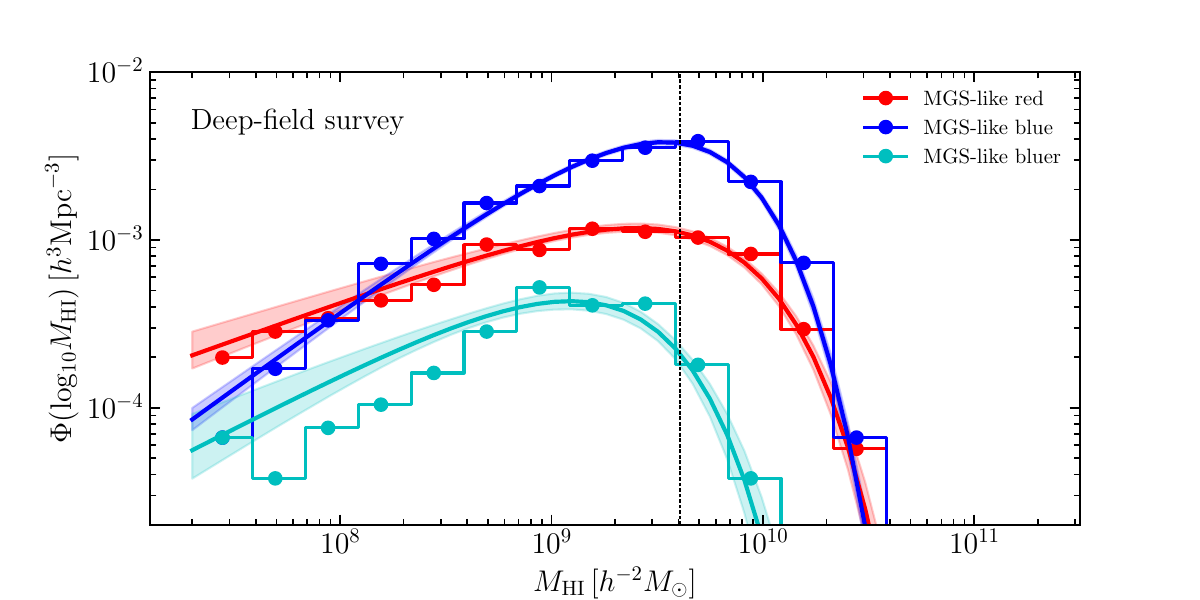}
  \includegraphics[width=0.5\textwidth]{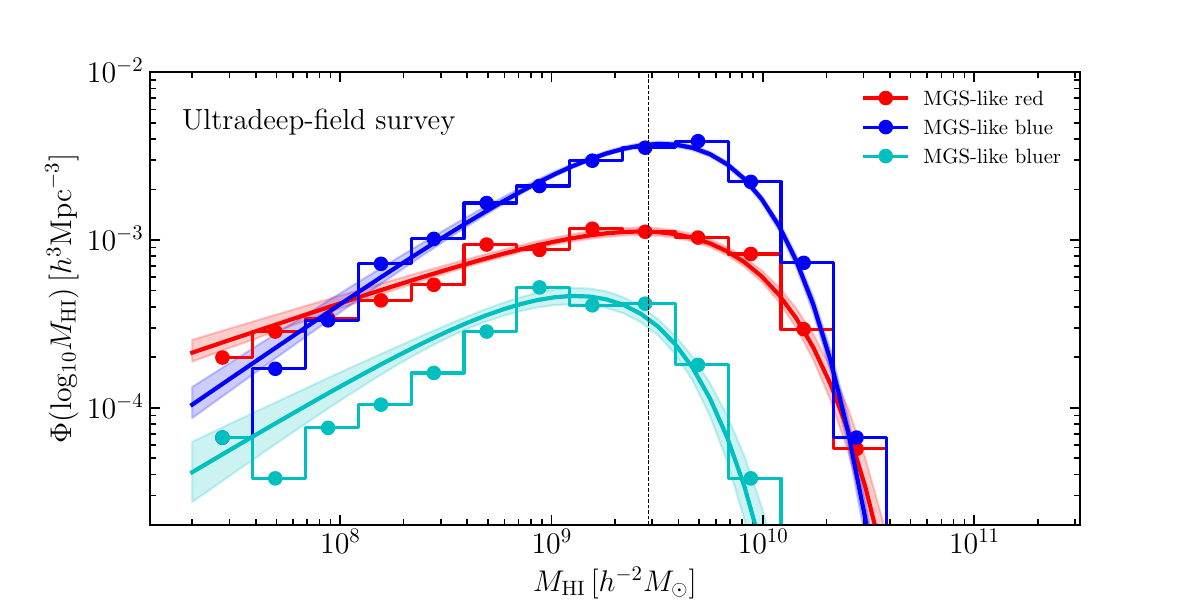}
\caption{The \hiMF estimated with the MGS-like sample. From top to bottom, different
panels show the results with the noise level of 
the pilot, deep-field, and ultradeep-field drift-scan observation, respectively (c.f. \reftb{tab:surveys}).
In each panel, the \hiMF estimated by Bayesian-stacking the red, blue, and bluer galaxy samples are 
shown with red, blue, and cyan colors, with the filled regions showing the $68\%$ confidence interval.
The dot-marked curves indicate the \hiMF estimated directly using the galaxy \hi mass 
extracted from the simulated catalog, i.e. $\Phi_{\rm g}(M_{\hi})$ as defined in \refsc{sec:himf}.
The vertical dashed lines indicate the \hi galaxy survey detection limit 
corresponding to $5$ times observation noise r.m.s..
\label{fig:HIMF_MGS}}
\end{figure}

Using the MGS-like galaxy catalog as the position proxy,  
we consider the \hiMF estimation by Bayesian-stacking the \hi flux from the simulation catalog.
The observation effect, e.g. the beam-smearing effect, is ignored, and 
the uncertainty raised from the measurement noise is emphasized. 
We simulate the measurements with the three different galaxy sub-samples, 
e.g. the red, blue, and bluer galaxy samples as defined in \refsc{sec:galaxypop}, individually. 
The simulation results are shown in \reffg{fig:HIMF_MGS}.

The results with different FAST noise levels are shown in different panels, 
i.e. from the top to the bottom, the results correspond to the noise level 
of the pilot, deep-field, and ultradeep-field drift-scan observation. 
In each panel, the \hiMF measurements, as well as the noise level, for the red, blue, 
and bluer galaxy samples are shown in red, blue, and cyan colors, respectively.

The vertical dashed line shows the $5$-times-noise-level equivalent \hi mass, 
estimated via \refeq{eq:mass2flux} assuming a mean redshift of $0.1$. 
We take $2\,{\rm MHz}$ as an approximation to the mean frequency width of different galaxy subsamples.

The five-times noise level represents a significant detection threshold for \hi galaxy survey 
and it is pushed to the low-mass end with repeating drift-scan observations.
Only galaxies that are bright in radio bands can be well cataloged via the
\hi galaxy survey.

The color-filled regions show the $68\%$ confidence intervals of the measurements.
We also present the \hiMF directly obtained with the \hi mass from 
the TNG simulation, i.e. $\Phi_{\rm g}$, using the dot-marked curves to examine the measurement accuracy.
The \hiMF measurements are consistent with the $\Phi_{\rm g}$ within the confidence intervals
for all three galaxy subsamples.
For each noise level injection, the Bayesian stacking method could significantly improve the
\hiMF measurements and constrain the \hiMF even below the 
$5$ times noise level, which estimates the detection limit for the \hi galaxy survey.

The FAST \hiim pilot and deep-field survey simulations assume a survey area of $210\,\deg^2$. 
The simulation results show that the \hiMF for both the red and blue galaxies can be well-constrained
with the FAST \hiim pilot survey. 
However, the \hiMF for bluer galaxies needs the deep-field survey, which is 4 times repeating 
observations of the pilot survey.

We assume a survey area of $210\,\deg^2$ for the FAST \hiim pilot, deep-field, and ultradeep-field 
surveys. The results for the pilot survey are shown in the top panel of \reffg{fig:HIMF_MGS}.
The simulation results show that the \hiMF for both the red and blue galaxies are 
well-constrained and consistent with $\Phi_{\rm g}$, while the bluer galaxies' \hiMF
has slightly larger measurement uncertainty due to the limited samples in the bluer galaxy catalog.

The results for a FAST deep-field and ultradeep-field survey are shown in the middle and bottom 
panel of \reffg{fig:HIMF_MGS}, respectively. 
The \hiMF for each of the galaxy catalogs is well-constrained.
Besides the bluer galaxy sample, the improvement in measurement uncertainty compared to the pilot survey 
is not significant.
This indicates that the measurement noise is subdominant, especially for the red and blue galaxy samples, 
and the errors on the \hiMF measurements are raised by the intrinsic scattering of the \hiMF 
between different halos, i.e. the sample variance \citep{2008IAUS..244..103S}.
The sample variance in \hiMF measurements may indicate an 
environment-dependent \hiMF, which requires conditional \hiMF estimation \citep{2022ApJ...941...48L}.

Instead of an ultradeep-field survey, a wide-field survey is more efficient
in beating down the sample variance. 
Generally speaking, with the same total observation time, a wide-field survey suffers 
from decreased sensitivity, which is tricky for traditional \hi galaxy surveys.
Beneficial from the Bayesian-stacking approach, the same sensitivity can be achieved
by stacking a large number of samples distributed across a shallow but wide survey field,
instead of an ultradeep one. Such a wide-field survey can be quickly carried out
with FAST drift-scan observation. 
For instance, if we assume that the galaxy number density is uniform across the sky, 
the noise level of the ultradeep-field survey is equivalent to stacking galaxy samples 
over a survey area of $\sim 1800\,\deg^2$.
Currently, limited by the TNG simulation box size, performing a wide-field
simulation is quite time-consuming. We will discuss the 
wide-field case further in future works. 

\begin{table}
\scriptsize
\begin{center}
\caption{The \hiMF parameter posterior distributions with MGS-like galaxy catalog. 
}\label{tab:paramMGS}
\hspace*{-1.0cm}
\renewcommand{\arraystretch}{1.2}
\begin{tabular}{lcccc} \hline\hline
                       & Pilot survey & Deep-field survey & Ultradeep-field survey \\ \hline
                       & \multicolumn{3}{c}{Red galaxy catalog} \\ \cline{2-4}
${\rm log}_{10}\Phi_*$ & $-2.99^{+0.05}_{-0.05}$         & $-2.99^{+0.04}_{-0.05}$              & $-3.00^{+0.03}_{-0.04}$ \\ 
${\rm log}_{10}M_*$    & $9.82^{+0.06}_{-0.05}$         & $9.81^{+0.05}_{-0.04}$              & $9.87^{+0.05}_{-0.04}$ \\ 
$\alpha$               & $-0.59^{+0.06}_{-0.06}$         & $-0.65^{+0.05}_{-0.08}$              & $-0.62^{+0.04}_{-0.05}$  \\ \hline
                       & \multicolumn{3}{c}{Blue galaxy catalog}    \\ \cline{2-4}
${\rm log}_{10}\Phi_*$ & $-2.35^{+0.02}_{-0.02}$         & $-2.35^{+0.01}_{-0.02}$              & $-2.38^{+0.02}_{-0.02}$ \\ 
${\rm log}_{10}M_*$     & $9.59^{+0.02}_{-0.02}$         & $9.57^{+0.02}_{-0.02}$              & $9.61^{+0.02}_{-0.02}$ \\ 
$\alpha$               & $-0.14^{+0.05}_{-0.05}$         & $-0.07^{+0.04}_{-0.04}$              & $-0.20^{+0.05}_{-0.06}$ \\ \hline
                       & \multicolumn{3}{c}{Bluer galaxy catalog}   \\ \cline{2-4}
${\rm log}_{10}\Phi_*$ & $-3.30^{+0.11}_{-0.15}$         & $-3.32^{+0.07}_{-0.08}$              & $-3.27^{+0.06}_{-0.07}$ \\ 
${\rm log}_{10}M_*$      & $9.27^{+0.14}_{-0.12}$          & $9.29^{+0.10}_{-0.08}$              & $9.23^{+0.09}_{-0.08}$ \\ 
$\alpha$                & $-0.28^{+0.28}_{-0.25}$          & $-0.40^{+0.14}_{-0.16}$              & $-0.22^{+0.15}_{-0.15}$ \\ \hline\hline
\end{tabular}
\end{center}
\end{table}

The constrain results on the \hiMF parameters for each case are listed in 
\reftb{tab:paramMGS}. 
All three parameters of red and blue populations can be well constrained with even a pilot survey. However, the parameters of the bluer galaxy catalog, which has limited galaxy samples, require a deep survey. We obtain a positive \hiMF faint-end slope index, $1 + \alpha$
for all the cases. 
Such \hiMF represents an optically selected magnitude-limited sample that lacks faint galaxies.
Indeed, when one gets closer to the mass resolution, the TNG simulation's sample 
completeness drops. The positive \hiMF faint-end slope index may indicate a
selection effect. Detailed discussions are shown in \refsc{sec:samplecomp} with full galaxy samples.

\subsection{Flux confusion effect for MGS-like catalog}

\begin{figure}
  \centering
  \includegraphics[width=0.49\textwidth]{plot_HIMF_MGS_cat_ultradeep.pdf}
  \includegraphics[width=0.49\textwidth]{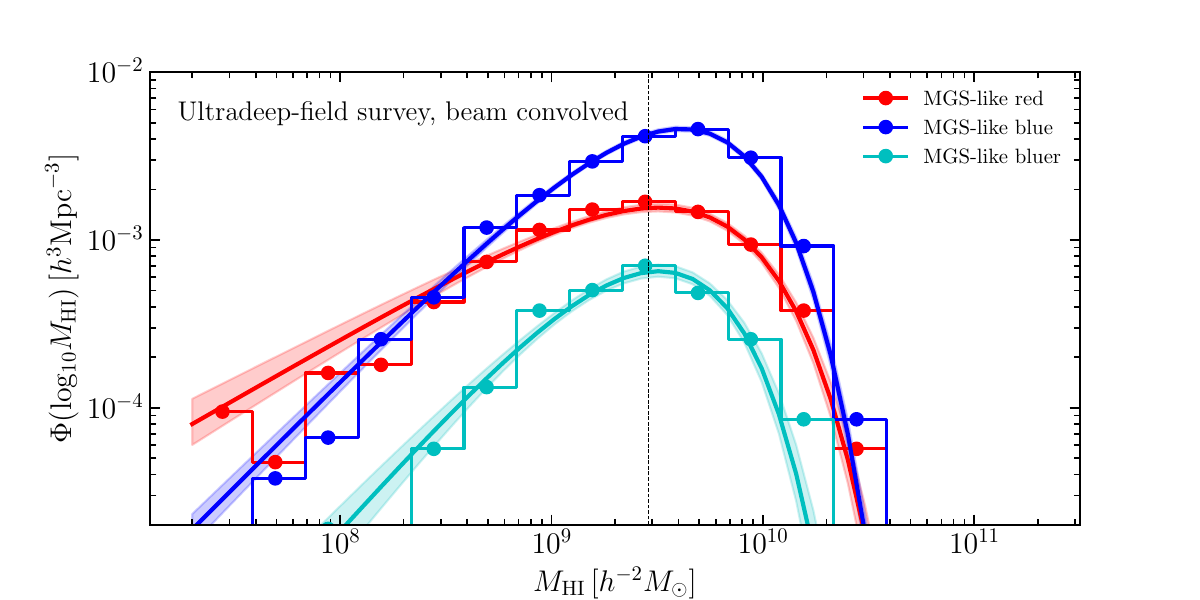}
  \includegraphics[width=0.49\textwidth]{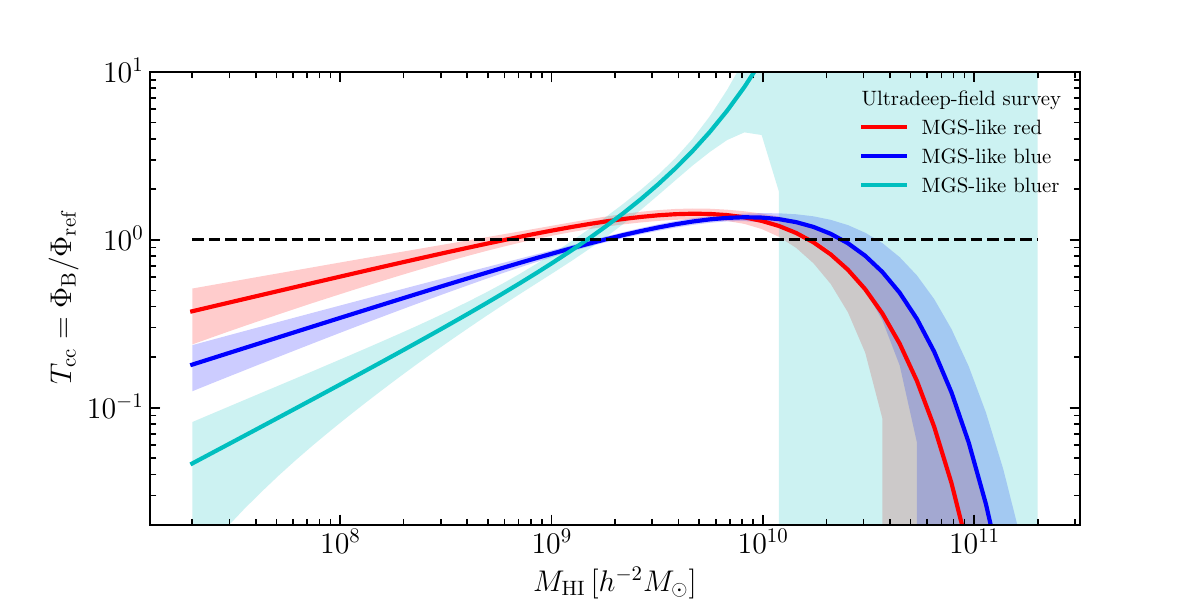}
\caption{The \hiMF measurements including the flux confusion effect.
The results with/without flux confusion effect are shown in the top/middle panels.
The bottom panel shows the confusion correction transfer function, $T_{\rm cc}$ (i.e. \refeq{eq:Tcc}).
The results for red, blue, and bluer galaxy catalogs are shown in red, blue, and cyan colors,
respectively. All the results assume a FAST \hiim ultradeep-field survey.
%wide-field survey.
\label{fig:HIMF_MGS_beam}}
\end{figure}

We further include the flux confusion effect using the MGS-like galaxy catalog
by convolving the simulated \hi sky cube with the FAST beam model.
The results with the confusion effect are shown in the middle panel of \reffg{fig:HIMF_MGS_beam}.
As a comparison, the top panel shows the original \hiMF estimation without the confusion effect.
The results are all with the assumption of the FAST \hiMF ultradeep-field survey to minimize the 
measurement uncertainty from the noise. 

The \hiMF for each galaxy sample is shifted towards the high-mass end because of 
the \hi flux of the galaxies within the beam resolution are mixed up.
We adopt the confusion correction transfer function, $T_{\rm cc}$ (i.e. \refeq{eq:Tcc}), 
to quantify the \hiMF shift. 
The transfer functions are displayed in the bottom panel of \reffg{fig:HIMF_MGS_beam} 
and are estimated for each galaxy catalog separately.
The errorbar for the transfer function is propagated from the measurements' uncertainty 
of the $\Phi_{\rm B}$  and  $\Phi_{\rm ref}$.
The transfer function is below $1$ at the low-mass end, where
the red galaxy sample exhibits a transfer function closer to $1$ compared 
to the blue and bluer galaxy samples, but the bluer galaxy shows a 
significant deviation from $1$.
It signifies a substantial influence of the confusion effects on \hiMF estimates, 
particularly for blue and bluer galaxies, which primarily consist of satellite galaxies. 
The transfer function increases with the galaxy \hi mass and eventually 
descends after a 'knee' mass, except for the bluer galaxy sample, which is noise-dominated 
at the high-mass end due to the absence of galaxy samples.

\subsection{The \hiMF improvement with FAST \hiim}

\begin{table}
\scriptsize
\begin{center}
\caption{The \hiMF parameter posterior distributions with MGS-like galaxy catalog compared with ALFALFA survey. 
}\label{tab:paramMGSvsALFALFA}
\hspace*{-1.0cm}
\begin{tabular}{lcccc} \hline\hline
                       & Deep-field survey & Ultradeep-field survey & ALFALFA$^\dag$ \\ \hline
                       & \multicolumn{3}{c}{Red galaxy catalog} \\ \cline{2-4}
$\sigma({\rm log}_{10}\Phi_*)$ & $ 0.045$         & $0.035$              & $0.055$ \\ 
$\sigma({\rm log}_{10}M_*)$    & $0.045$         & $0.045$              & $0.040$ \\ 
$\sigma(\alpha)$               & $0.065$         & $0.045$              & $0.100$  \\ \hline
                       & \multicolumn{3}{c}{Blue galaxy catalog}    \\ \cline{2-4}
$\sigma({\rm log}_{10}\Phi_*)$ & $0.015$         & $0.020$              & $0.029$ \\ 
$\sigma({\rm log}_{10}M_*)$    & $0.020$         & $0.020$              & $0.030$ \\ 
$\sigma(\alpha)$               & $0.040$         & $0.055$              & $0.110$ \\ \hline
                       & \multicolumn{3}{c}{Bluer galaxy catalog}   \\ \cline{2-4}
$\sigma({\rm log}_{10}\Phi_*)$ & $0.075$         & $0.065$              & $0.089$ \\ 
$\sigma({\rm log}_{10}M_*)$    & $0.090$         & $0.085$              & $0.050$ \\ 
$\sigma(\alpha)$               & $0.155$         & $0.150$              & $0.380$ \\ \hline\hline
\end{tabular}
\end{center}
{\footnotesize
\noindent
$^{\dag}$ Parameter constraint results are from \citealt{2020MNRAS.494.2664D}
}
\end{table}

\begin{figure}
  \centering                           
  \includegraphics[width=0.5\textwidth]{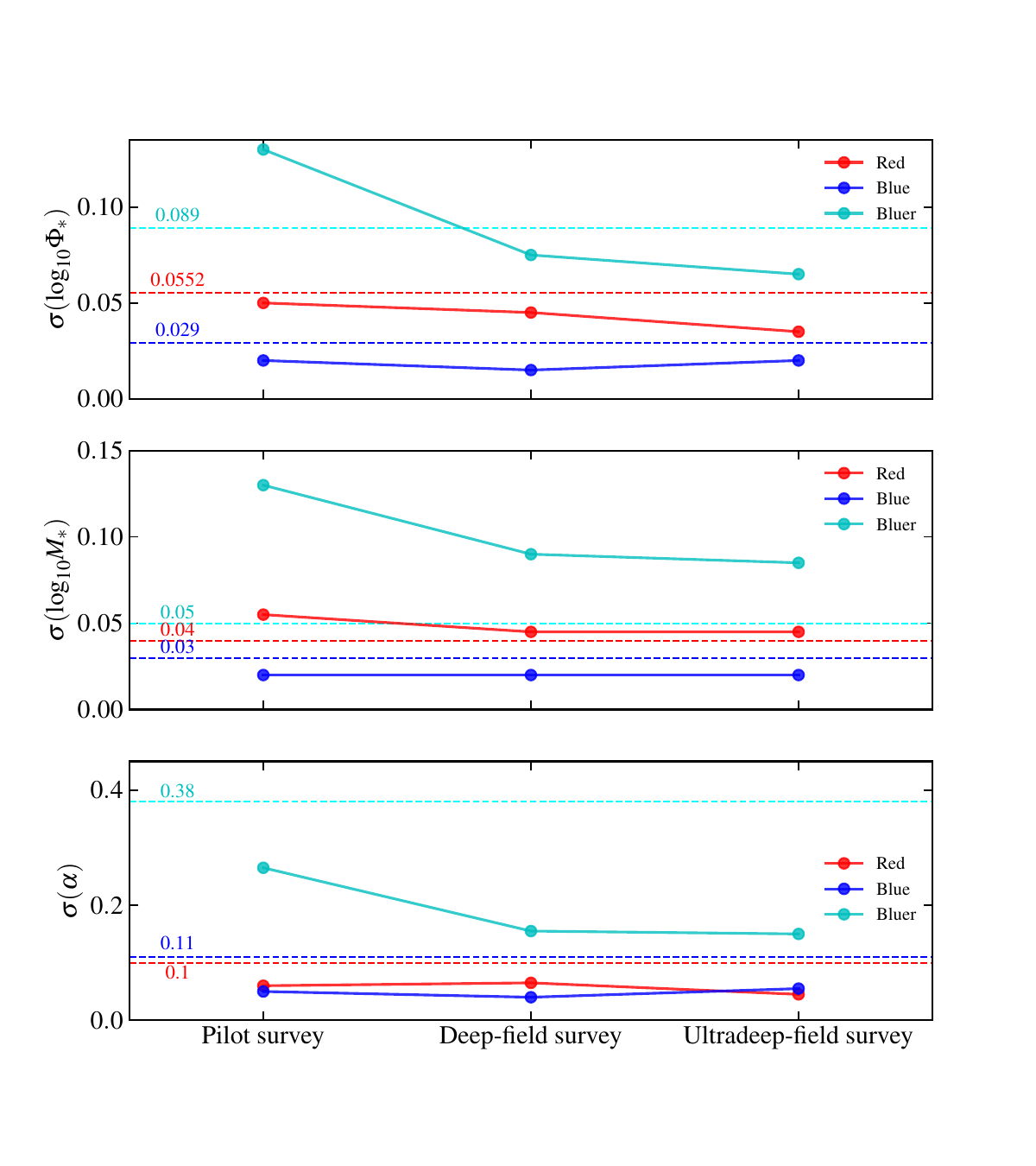}
\caption{
The \hiMF parameter 1 $\sigma$ errors of MGS-like galaxy catalog compared with ALFALFA survey.
\label{fig:paramerror}}
\end{figure}

The constraint errors of the \hiMF parameters with different FAST \hiim survey assumptions 
are listed in \reftb{tab:paramMGSvsALFALFA}. 
The table contains 1 $\sigma$ \hiMF parameter errors for red, blue, and bluer galaxy catalog
from top to bottom.
We list the errors of the deep-field survey (the left column),  
the ultradeep-field survey (the middle column), and the corresponding constraint errors with
ALFALFA survey from \citet{2020MNRAS.494.2664D} (the right column), respectively. 
In addition, in order to more intuitively observe the change of error with the 
integration time and compare the results with ALFALFA survey, 
we plot $1\sigma$ errors of three main \hiMF parameters in Fig.~\ref{fig:paramerror}. 
The horizontal coordinate represents the FAST pilot survey, deep-field survey, 
and ultradeep-field survey, respectively. 
The constraint errors of the red, blue, and bluer galaxies are shown in red, blue, and cyan,
respectively.
The dashed horizontal lines indicate the 1 $\sigma$ errors of the \hiMF parameters obtained by \citet{2020MNRAS.494.2664D}.
Benefit from the high sensitivity of FAST, the constraint error with 210 ${\rm deg}^2$
of the pilot survey is compatible with those obtained by the ALFALFA survey 
with nearly 7000 ${\rm deg}^2$.

The constraint errors gradually decrease with deeper \hiim surveys, however, 
the error reduction is not significant.
As mentioned above, we infer that the error of the sample is dominated by the sample variance, 
instead of the observational thermal noise. 
Due to the limited samples of the bluer galaxies, the constraint errors on $M_*$
is generally greater than the current ALFALFA results.
Therefore, rather than extending the integration time, 
a large area survey would be ideal for stacking-based \hiMF measurements.

\subsection{Selection completeness}\label{sec:samplecomp}

\begin{figure}
  \centering
  \includegraphics[width=0.49\textwidth]{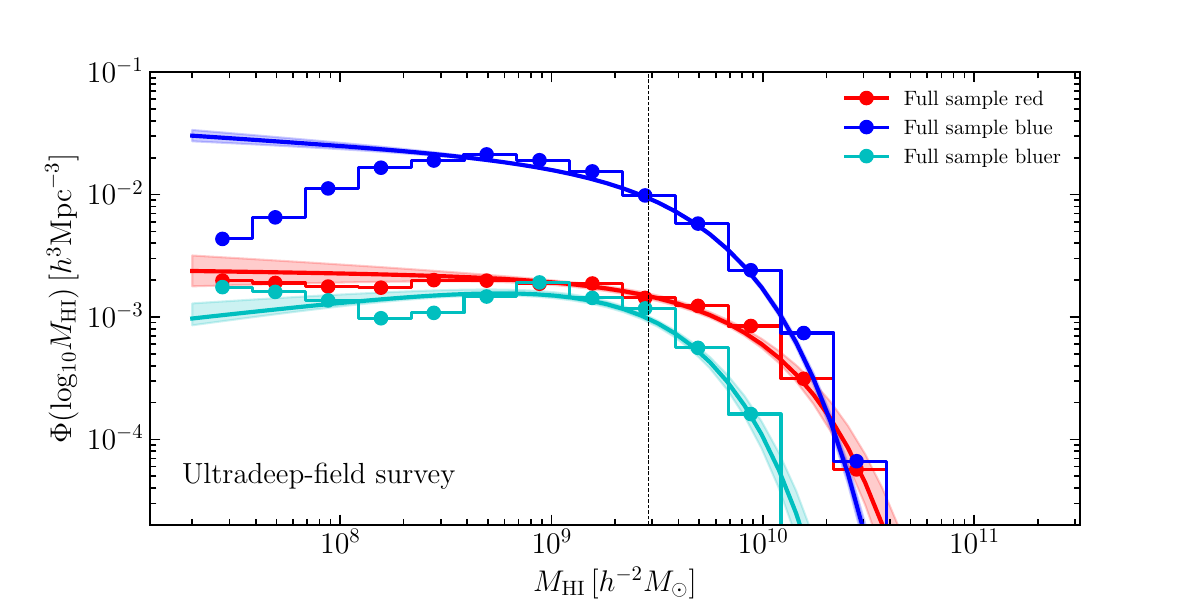}
  \includegraphics[width=0.49\textwidth]{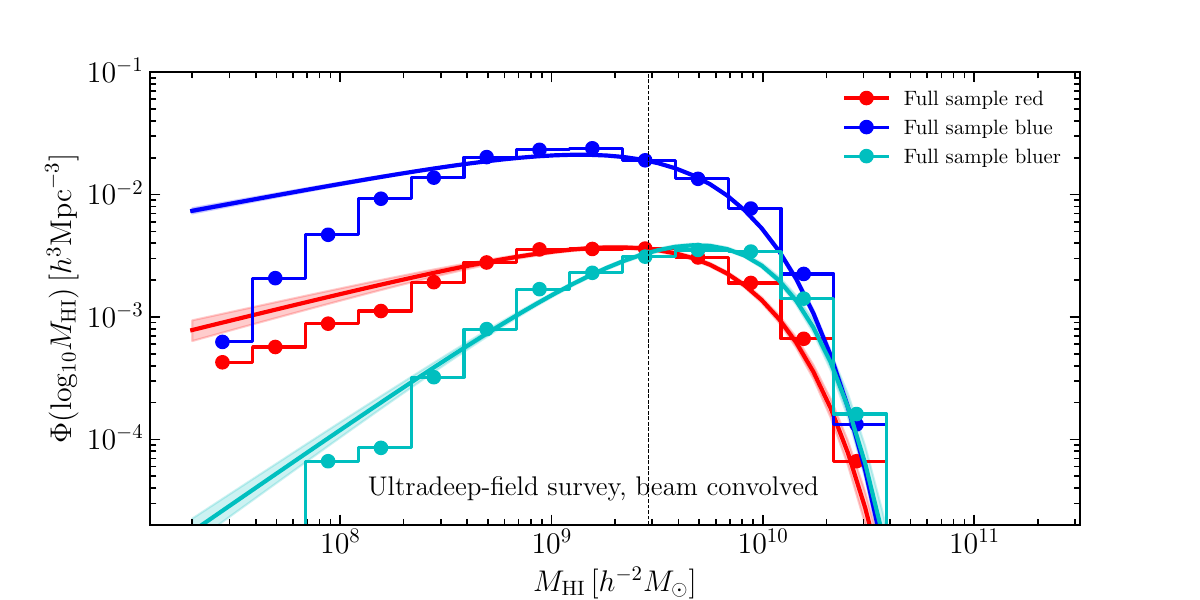}
  \includegraphics[width=0.49\textwidth]{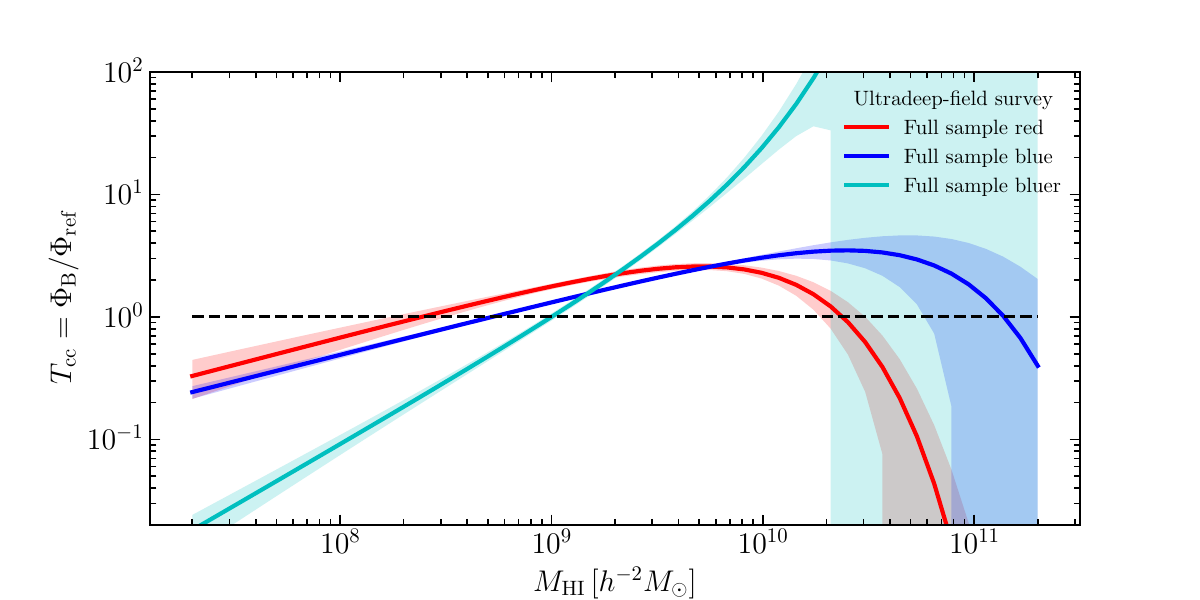}
\caption{Same as \reffg{fig:HIMF_MGS_beam} but for the full galaxy sample.
\label{fig:HIMF_full_sample_cat}}
\end{figure}

The \hiMF estimation is generally based on a volume-limited sample. 
Our simulation sky box is constructed with a snapshot and the
initial galaxy sample is close to a volume-limited sample as 
the magnitude limit variation is negligible within the simulated 
comoving distance range. 

We construct a full galaxy sample with all the galaxies identified 
in the TNG simulation snapshot and split it into red, blue, and bluer
galaxy samples following the same partition method described in 
\refsc{sec:galaxypop}. 
Using the full galaxy sample and assuming with the ultradeep-field survey, 
we apply the stacking-based \hiMF estimation 
to the \hi cube without and with the beam confusion effect, respectively.
The results are shown in \reffg{fig:HIMF_full_sample_cat}, where
the top and middle panels show the results without and with beam
confusion effect, respectively.

Without the beam confusion effect, the \hiMF generally fits well with 
the Schechter function, except for the blue galaxy sample, which 
is significantly deviated from the Schechter function at the lower end
of the \hiMF. Such deviation may indicate an intrinsic incompleteness
of the TNG galaxy sample with the galaxy's stellar mass approaching the mass resolution of the simulation.

With the confusion effect, we observe a similar effect as the MGS, i.e. 
the \hiMF is shifted towards the high-mass end. The confusion correction 
function for the full galaxy samples is shown in the bottom 
panel of \reffg{fig:HIMF_full_sample_cat}.

\begin{figure}
  \centering
  \includegraphics[width=0.49\textwidth]{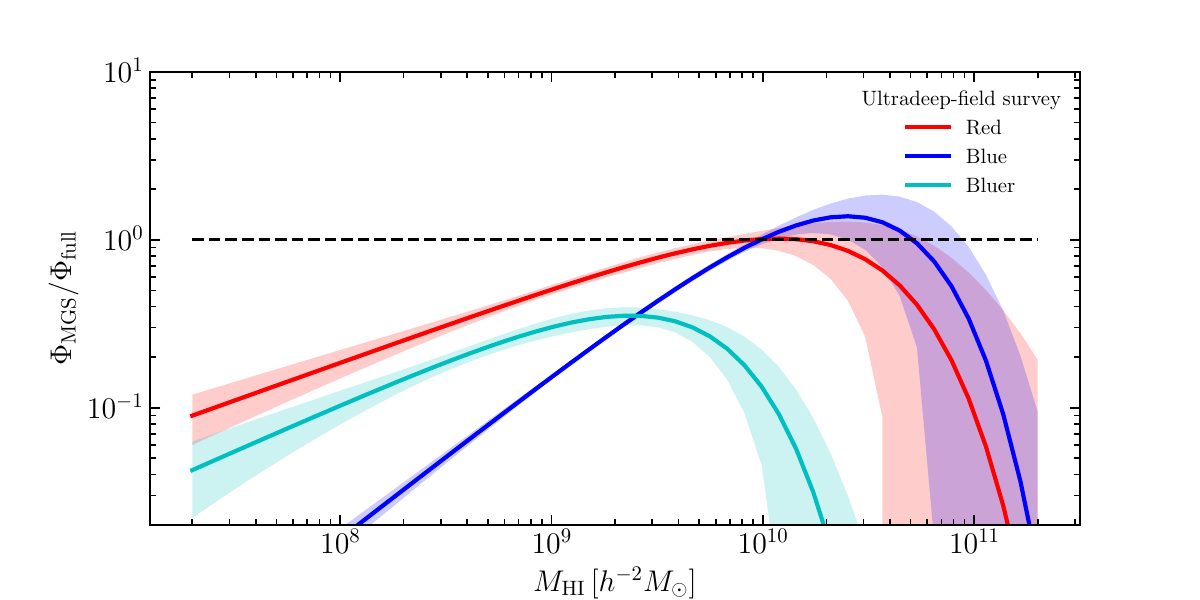}
\caption{
The \hiMF ratio of the MGS-like sample with respect to the full sample, assuming the ultradeep-field drift-scan
observation. 
The results with the luminous red, blue, and bluer galaxy catalog are shown in red, blue, and cyan colors, respectively.
\label{fig:HIMF_full_sample_ratio}}
\end{figure}

However, the stacking-based \hiMF estimation is also affected by the 
completeness of the optically selected sample, in particular, the 
MGS-like galaxy sample,
which is known to be a magnitude-limited sample \citep{2002AJ....124.1810S}.
We estimate the \hiMF ratio of the MGS-like galaxy samples with
respect to the full galaxy sample. The results are shown in 
\reffg{fig:HIMF_full_sample_ratio}. 
The red, blue, and cyan colors represent the luminous red, blue, and bluer 
galaxies, respectively. 
The error represents the estimation uncertainty propagated from the measurement
uncertainty of the \hiMF. 
It is important to note that the accuracy of the transfer function decreases 
at the higher \hi mass end due to the smaller number of galaxies available in the sample.
Both the red and blue galaxy subsamples exhibit transfer function values below unity 
at the lower \hi mass end, gradually approaching unity with \hi mass increases, and 
dropped at the higher-mass end. 
In contrast, the bluer galaxy subsample shows an increase in the transfer function 
with \hi mass but remains consistently below unity, even at the higher mass end.
The deviation from unity indicates that there are galaxies 
are excluded with the optical selection magnitude limit, even with large \hi mass.

Notably, at the low \hi mass end, the blue galaxy subsample deviates more significantly 
from unity compared to the red galaxy subsample. This suggests that blue galaxies 
with lower \hi masses are less completely represented in the optically selected sample 
than red galaxies of comparable optical brightness.

The impact of sample completeness on the estimation of the \hiMF has been discussed 
extensively in the literature. For example, in a series of studies on \hiMF estimation 
using data from the ALFALFA \hi galaxy survey, the effects of sample completeness 
have been analyzed \citep{2021MNRAS.500L..37D,2022MNRAS.511.2585D}. 
Additionally, \citet{2022ApJ...941...48L} proposed an \hi mass estimator that incorporates 
various galaxy properties to provide unbiased \hi mass estimates.

However, we emphasize that the sample completeness addressed in this work 
is fundamentally different from that considered in previous studies. 
Here, the completeness issue arises from the magnitude limit in the optical band, 
rather than the flux limit in radio observations. 
This distinction is particularly significant for Bayesian stacking-based \hiMF estimation, 
which relies on optically selected galaxy samples.
Since establishing strategies to limit the influence of optical selection effects 
is beyond the scope of this study, we defer this task to future research to 
enhance \hiMF calculations.

\section{Conclusion}\label{sec:conc}

In this work, we simulate the Bayesian-stacking-based \hiMF estimation with FAST \hiim drift scan survey. 
Using one of the simulation snapshots of the TNG50, we simulated the \hi cube, as well as the corresponding 
optical galaxy survey catalog, at redshift of $z\sim0.1$, which aligns with one of the RFI-free frequency
band of FAST L-band.

Following the previous \hiMF analysis \citep{2020MNRAS.494.2664D}, we simulate galaxy catalog according to
the galaxy selection criteria of the main galaxy sample (MGS) of SDSS Data Release 7, and split the 
MGS-like catalog into three subsamples, i.e. the red, blue, and bluer galaxy catalogs, based on
the galaxy characteristics in the color-magnitude plane.

We also consider the observational thermal noise and the confusion effect caused by the FAST beam.
Assuming the survey area of $210\,{\rm deg}^2$, we simulate the \hiim survey with different 
observation integration time per beam, i.e., $29\,{\rm s}$ as the Pilot survey, 
$115\,{\rm s}$ as the Deep-field survey', and $230\,{\rm s}$ as the Ultradeep-field survey.

The \hiMF can be effectively reconstructed using Bayesian-stacking-based \hiMF estimation 
for the red, blue, and bluer galaxy catalogs, utilizing the observation integration times 
of the Pilot survey, Deep-field survey, and Ultradeep-field survey.
Due to the limited sample amount, the \hiMF reconstruction error for bluer galaxy catalog
is larger than the other two galaxy catalogs.
The improvement of the constraint uncertainty with more integration time is not significant,
which indicates currently \hiMF estimation is sample variance dominated.
A large field \hiim survey is efficient for reducing the sample variance and robust for 
Bayesian-stacking-based \hiMF estimation.

We also find a significant confusion effect due to the limited angular resolution of the FAST beam.
The confusion effect shifts the \hiMF to the higher-mass end and thus may cause underestimation 
of the \hiMF at the lower-mass end. With the simulation, we adopt a transfer function to 
characterize and correct the confusion effect for the red, blue, and bluer catalogs, respectively.

The effect caused by the sample incompleteness is notable in our simulation.
An intrinsic incompleteness arises from the TNG galaxy sample when the stellar mass of the galaxies approaches 
the mass resolution limit.
Besides, the Bayesian-stacking-based \hiMF estimation is also affected by the completeness of the
optically selected sample. With our simulation, we imply a transfer function to characterize the 
completeness effect for the MGS-like catalog for the red, blue, and bluer catalogs, respectively. 

With this work, we highlight a few systematic effects for Bayesian-stacking-based \hiMF estimation for 
current and future \hiim surveys with FAST. 
This study serves as a fundamental preparation for further \hiMF analysis with our 
FAST drift scan surveys for \hiim. 

\section*{acknowledgments}

We acknowledge the support of the National SKA Program of China 
(Nos.~2022SKA0110200, 2022SKA0110203, 2022SKA0110100, 2022SKA0110101),
the National Natural Science Foundation of China (Nos. 12473091, 12473001,12361141814), 
111 Project (No. B16009), and the NSFC International (Regional) Cooperation and Exchange Project (No. 12361141814).
We express our gratitude to ChatGPT for its assistance in refining the manuscript and enhancing the code to improve execution efficiency.

\section*{Data Availability}

The corresponding author is available to share the simulated data underlying this article upon reasonable request.

\software{{\sc Astropy} \citep{2013A&A...558A..33A,2018AJ....156..123A,2022ApJ...935..167A},  
          {\sc Corner} \citep{corner},
          {\sc MultiNest} \citep{2009MNRAS.398.1601F}
          }

%\appendix
%\section{Appendix information}

\bibliography{main}{}
\bibliographystyle{aasjournal}



\end{CJK*}
\end{document}